\documentclass[english,pre,twocolumn,nofootinbib,notitlepage]{revtex4-1}

\usepackage[paper=letterpaper,margin=0.75in]{geometry}
\pdfpagewidth=\paperwidth
\pdfpageheight=\paperheight

\usepackage[T1]{fontenc}
\usepackage[utf8]{inputenc} 

\usepackage{natbib}
\usepackage[english]{babel}
\usepackage{bm,booktabs,amsmath,mathtools}
\usepackage{graphicx,array,float}
\usepackage{csquotes}
\usepackage[normalem]{ulem}
\usepackage{subcaption} 

\usepackage[dvipsnames]{xcolor}
\usepackage[colorlinks=true,linkcolor=black,urlcolor=black,
            citecolor=MidnightBlue,anchorcolor=MidnightBlue]{hyperref}

\newcolumntype{C}[1]{>{\centering\arraybackslash}p{#1}}


\begin{document}

\title{Intracellular phagosome shell is rigid enough to transfer outside torque to the inner spherical particle} 
\author{Srestha Roy$^{1,*}$, Arvin Gopal Subramaniam$^{1,*}$, Snigdhadev Chakraborty$^{1}$, Jayesh Goswami$^{1}$,Subastri Ariraman $^{2}$, Krishna Kumari Swain$^{1}$, Swathi Sudhakar$^{2}$,  Rajesh Singh$^{1,\ddagger}$, and Basudev Roy$^{1,\ddagger}$}
\affiliation{%
$^1$Department of Physics, IIT Madras, Chennai 600036, India
}
\affiliation{%
$^2$Department of Applied Mechanics, IIT Madras, Chennai 600036, India
}%

\begin{abstract}
\section*{Abstract}
Intracellular phagosomes have a lipid bilayer encapsulated fluidic shell outside the particle, on the outer side of which, molecular motors are attached. An optically trapped spherical birefringent particle phagosome provides an ideal platform to probe fluidity of the shell, as the inner particle is optically confined both in translation and in rotation.  Using a recently reported method to calibrate the translation and pitch rotations - yielding a spatial resolution of about 2 nm and angular resolution of 0.1 degrees - we report novel roto-translational coupled dynamics. We also suggest a new technique where we explore the correlation between the translation and pitch rotation to study extent of activity. Given that a spherical birefringent particle phagosome is almost a sphere, the fact that it turns due to the activity of the motors is not obvious, even implying high rigidity of shell. Applying a minimal model for the roto-translational coupling, we further show that this coupling manifests itself as sustained fluxes in phase space, a signature of broken detailed balance.
\end{abstract}

\maketitle
\def\thefootnote{$\ast$}\footnotetext{Equal contributions}
\def\thefootnote{$\ddagger$}\footnotetext{basudev@iitm.ac.in, rsingh@iitm.ac.in}
\section*{Introduction}

Endocytic pathways are crucial for the cell. It is through these that the cell "eats" and "drinks", to nourish itself and to remove pathogens from bodies. There can be three different types of mechanisms for this pathway, namely, pinocytosis, endocytosis and phagocytosis. Of these, phagocytosis corresponds to the case when the object to be eaten is larger than about 500 nm\cite{aderem1999mechanisms,baranov2021modulation}. \textcolor{black}{The engulfed phagosomes }that appear from the process of phagocytosis are also important from a biophysical study standpoint as these constitute the tracers which can then be tracked for single particle tracking, rheology, study of transport problems and so on\cite{lau2003microrheology,lee2006ballistic,lee2007nuclear,betz2024characterizing,muenker2024accessing}. However, it is generally believed that phagosomes have an outer membrane consisting of the very lipid bilayer that is around the cell, and has extracellular fluids between the particle and the membrane. \textcolor{black}{It is also believed that the molecular motors outside the phagosome slide on the membrane diffusively. Thus, the shell is believed to be very fluid-like. This begs the question: is this shell indeed so fluid that it cannot transfer torque from outside the phagosome to the interior particle?} We address this question in this paper.

Several approaches have been investigated 
to measure active mechanisms due to molecular motors \cite{arbore2019probing,blehm2013vivo,rief2000myosin,veigel2005load,cappello2007myosin,ArneGennerich2014improved,block2003probing,jeney2004mechanical,ahmed2018active}. Plus ended single kinesin motors are capable of exerting an active force of about 5 pN whereas minus ended dynein can exert around 1.1 pN as an individual motor. Intracellular active forces contributed by molecular motors are the driving force behind emergent functionality observed at organ level \cite{haertter2024stochastic,hirokawa2010molecular}.  Several neurodegenerative diseases stem from disrupted movement of cargo\cite{deVos2008role,millecamps2013axonal}. Some studies suggest that long distance transport is brought about by coordination between multiple motors forming a team, whereas others claim dominance of a single active motor at any instant of time\cite{vershinintauregulation}. 

Transport of cargo along microtubules is directional, though complexity due to 3-D structure of cytoskeletal network and intracellular crowding poses hurdles in the path. The nature of motion largely depends on the local cytoskeletal architecture through which they pass\cite{bergman2018cargo}. Interesting dynamics are observed as motor carried cargo maneuvers around ``roadblocks" often caused by intersecting microfilaments\cite{ross2008cargo,gao2018cargos}. Motion around microtubule junctions are often marked by long pauses
\cite{balint2013correlative, verdeny20173d,mudrakola2009optically,kapitein2010mixed}, deformation of transported endosomes\cite{zajac2013local}. Most of the exploratory studies on motor dynamics are in vitro owing to difficulties in tracking in vivo motion with high resolution. These studies are further limited by inability to mimic the intracellular environment completely. 

Here, we explore both the translational motion of a phagosome and the out-of-plane (or pitch, as shown in Fig.\ref{schematic}(a)) rotation simultaneously at high resolution using photonic force microscopy, while the cargo is trapped in the linearly polarized optical tweezers. We ask whether a correlated rotational motion is observed while the phagosome exhibits translational motion. Given that the optical tweezers applies a restoring torque to align the particle towards the polarization axis, even in the pitch sense\cite{friese1998optical,roy2016directed}, any observed rotation would be a signature that the torque applied by the molecular motors on the outside of the phagosome does indeed get transferred to the internal spherical birefringent particle. This in turn implies that the shell of the phagosome is quite rigid. This paves the way for simultaneous detection of rotation with optical tweezers alongside translation to explore intracellular transport. We also find that directed three-dimensional motion inside the cell tends to turn the particle in the out-of-plane degree of rotational freedom via selected distinct mechanisms. 

The results become important because the tracer is now spherical in shape with less complications than due to the anisotropic drag while using the rod or anisotropic shaped particles conventionally used as reported in \cite{chen2017characteristic,kaplan2018rotation,song2024deep}. 

The presence of molecular motors on the phagosomal exterior surface is reasonably well established. It has been shown that dynein is quite uniformly distributed on phagosomes by immunofluorescence \cite{rai2016dynein}. Endogenous kinesin is difficult to detect, but can also be assumed to be uniformly distributed on such phagosomes. It is also known that several proteins diffuse within a lipid bilayer with diffusion constant of about D = 10 $\mu$m$^2$/s \cite{weiss2013quantifying}. For kinesin, this constant was reported to be  about D = 1.4 $\mu$m$^2$/s \cite{grover2016transport}. The time spent by a diffusing protein within the contact area between a microtubule filament and the phagosome, T, is about 0.08/(4 $\times$ D) \cite{sanghavi2018coin}. Depending on these above values of D, T ranges from 2 to 14 ms. However, reversals, stalls and motion \cite{sanghavi2018coin} of a phagosome inside a cell while being optically confined usually extends over 1 sec or more, so that diffusion of motors (both into and out of the contact area) happens much faster. Thus, the forces measured are unaffected by lateral diffusion of motors on the phagosome, if indeed such diffusion happens and mainly averaged in nature. We ask the question whether the averaged torques due to such forces are also unaffected due to such motion. 

Having established that the pitch rotations and translation degrees of freedom are coupled, we then propose a minimal model that captures this coupling, and further show how this model displays non-equilibrium signatures (rendered by the underlying molecular motors in the cytoskeleton), which we quantify via phase space fluxes and steady-state entropy production \cite{gnesotto2018broken, battle2016broken}.

\section*{Results\label{sec:results}}

\begin{figure}
    \includegraphics[width=\linewidth]{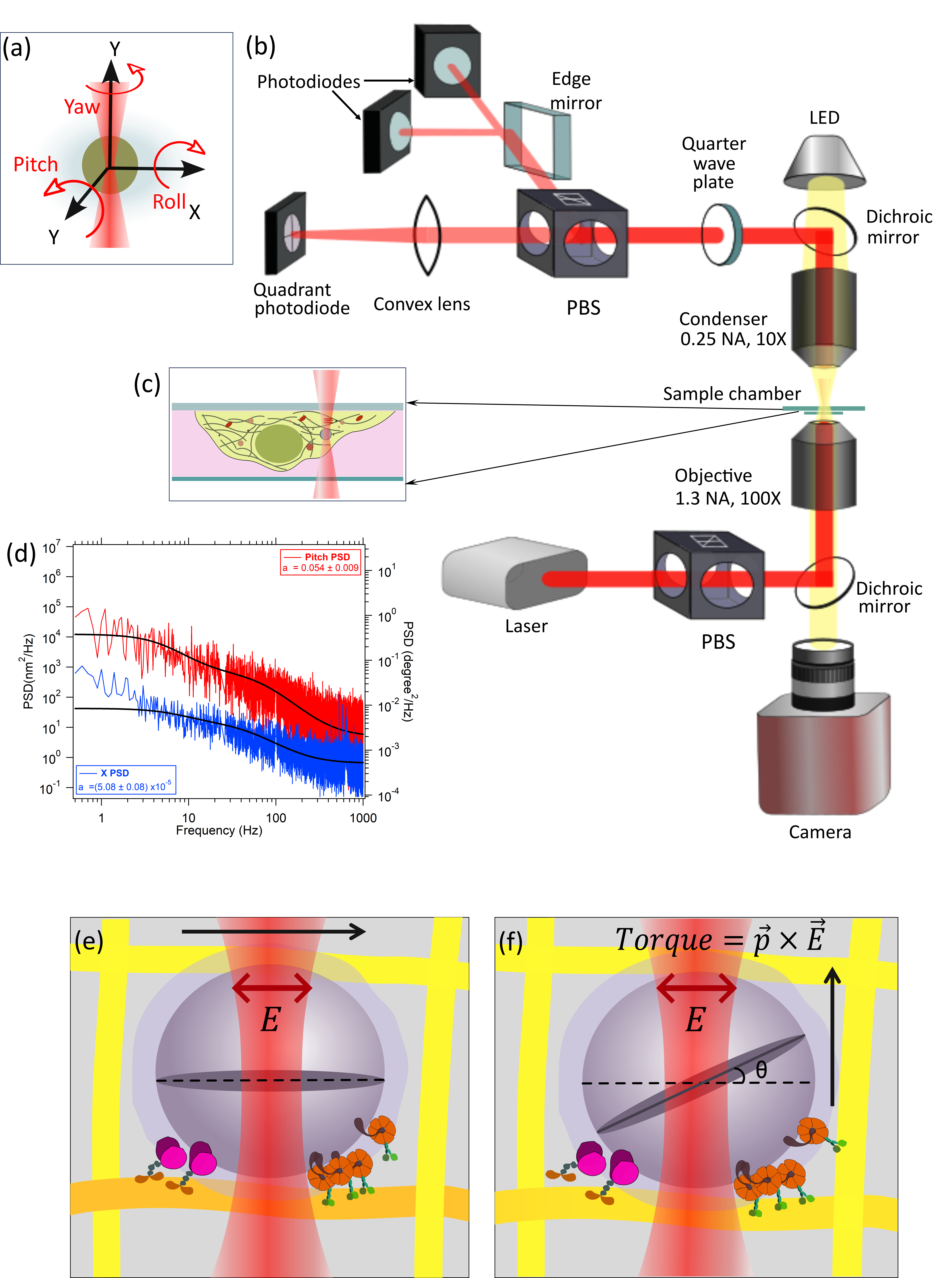}
    \caption{Schematic for study of pitch rotation adn translation of phagosomes inside cell
        (a) The six degrees of freedom of an optically trapped sphere.
    (b) Schematic of experimental setup.
    (c) The sample chamber will cells attached to the top surface and are maintained within a layer of DMEM media. Phagocytosed beads are trapped with the focused beam.
    (d) The power spectral densities (PSDs) for X translation and pitch rotation of a phagocytosed bead in a cell fitted to eq. Obtained fitting parameter is used to calibrate the PSD. 
    Panels (e) and (f) show oppositely directed motors pulling on a birefringent cargo. 
    The darker shaded region indicate the plane of optic axis. Detachment of some motors and new attachments formed in (f) causes the cargo to move forward and rotate, shown by rotation of optic axis. }
    \label{schematic}
\end{figure}

In order to study the problem, we study phagocytosed birefringent beads(PBB) made out of a liquid crystalline material. We choose a cell with a PBB that is far from the nucleus. The stage is moved to bring the particle close to the trap position. A typical schematic diagram of the experiment and the question we are asking has been depicted in Fig. \ref{schematic}(b).

The focused light is illuminated on the PBB to confine both the translation and rotation. Here, we note that the local trap stiffness is known to vary inside the cell from region to region owing to heterogeneity of cytoplasmic refractive index. Moreover, the exact refractive index of the regions are unknown. Thus, it has been difficult to calibrate the optical tweezers inside the cell. We use a method introduced in \cite{nakul2023studying} to calibrate the motion and estimate localized trap stiffness. We go ahead to study the motion of the PBB while simultaneously tracking both the translation and rotation. 

\subsection*{Correlated motion: Presence of pitch rotations during intracellular motion}

Particles phagocytosed into the cells may be bound to several molecular motors, which move them on the cytoskeletal tracks. The cytoskeletal tracks form an entangled mesh-like 3-D network (see Fig. \ref{schematic}(c), (e) and (f)) inside the cell\cite{ahmed2018active}. Therefore, a PBB inside a cell is pulled in different directions by motors bound to these fibers, a phenomenon known as "active diffusion" which is known to accelerate diffusive processes. There are many kinds of molecular motors inside the cell. Kinesin and dynein acting on microtubule filaments walk in opposite directions. Kinesin steps are directed towards the + end and dynein steps are towards the - end of microtubule and are referred to as anterograde and retrograde motion respectively. Myosin motors walk on actin filaments towards + end\cite{mukherjee2013electrostatic}. 

\begin{figure*}
    \centering
    \includegraphics[width=0.95\linewidth]{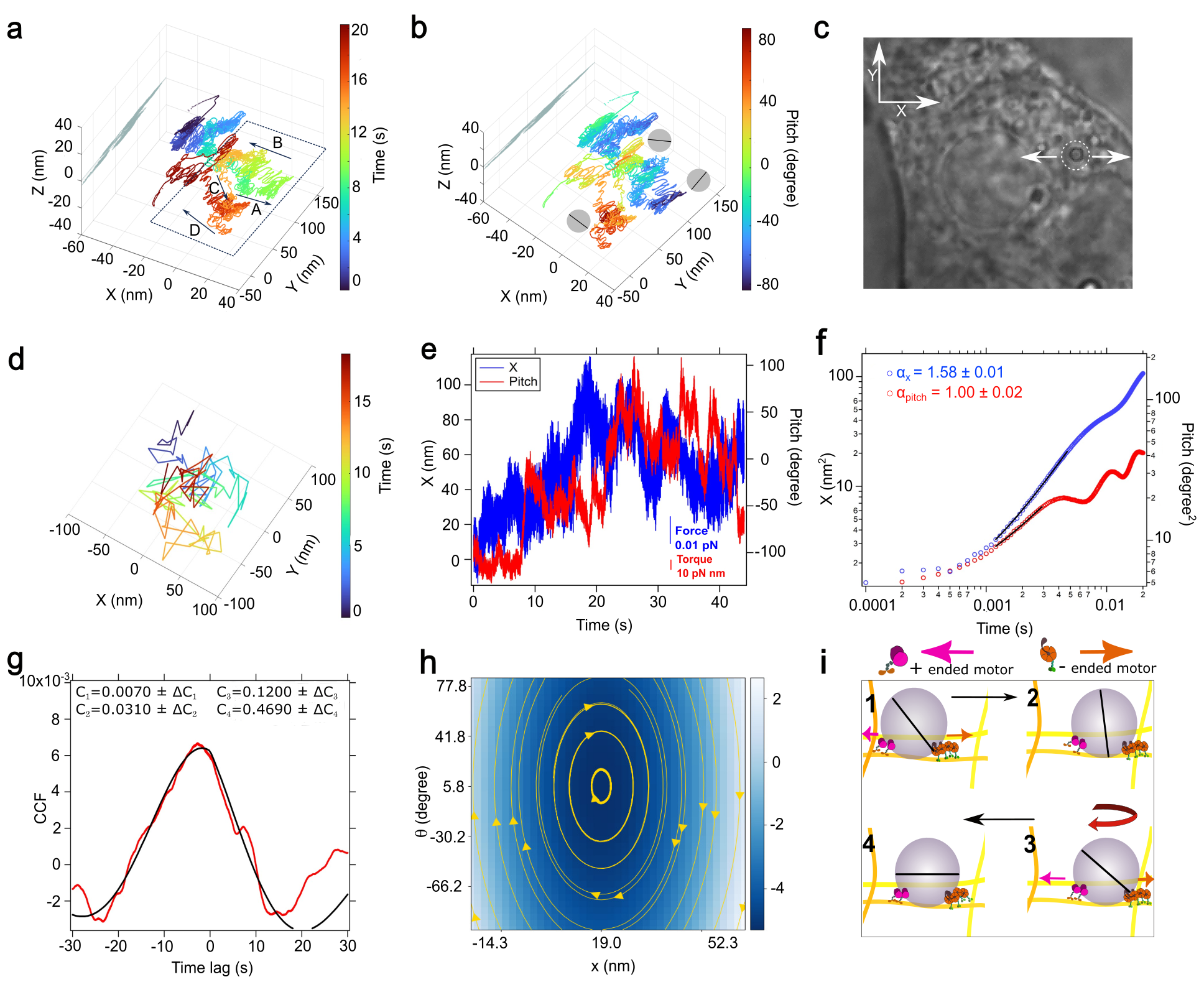}
    \caption{Tug-of-war like motion demonstrated by phagosome. 4D trajectory of phagosome with (a) time shown in colour,
    (b) pitch angle shown in colour, (c) position of the bead inside cell captured with a bright field microscope. White arrows indicate the direction of to and fro motion (d) 2D trajectory in X-Y plane extracted from video tracking.
    (e) Time series for X and pitch for tracked with QPD with respective force and torque scale bars.
    (f) Corresponding MSD curves for X (blue open circle) and pitch (red open circle). (g) CCF of X and pitch rotation from experimental data (red curve) and fit to Eq.\ref{eq:CCF} (black curve). (h) Broken detailed balance is demonstrated by showing circulating current from the vector plot of the configuration current, see Eq.\eqref{eq:current}. Using the parameters obtained by fit to the data from (g), we do a vector plot of the configuration current, see Eq.\eqref{eq:current}, which is overlaid on the pseudo-color plot of the steady-state probability distribution, see Eq.\eqref{eq:probS}.
    (i) Oppositely directed teams of motors pulling on a cargo along a cytoskeletal track causes the birefringent cargo to rotate. The optics axis shown as the black solid line indicates the orientation of the birefringent sphere. Right end directed team of motors pulls on it causing it to rotate in clockwise sense (1-2). The other team of motors take over to cause reversal in direction and rotation in anticlockwise direction (2-3 and 3-4). }
    \label{tugofwar}
\end{figure*}

\subsection*{Distinct classes of motion}
We experimentally measure the roto-translational dynamics observed during intracellular transport. We classify the types of motion into three broad categories: (i) 
rotation during tug-of-war,
(ii) rotation during unidirectional pulling of cargo,
and (iii) rotation while overcoming obstacles in the path during translational motion. For each class of motion, we measure and compare the following: (a) 4D phase space dynamics (for time and pitch angle separately), (b) roto-translational correlation, (c) phase-space roto-translational fluxes and steady-state entropy production rate. The latter two quantities are defined and explained in detail in Section \ref{sec:theory} later.

\subsection*{Tug-of-war}

When oppositely directed motors exert similar forces on the same cargo, a competition arises between the two teams which is termed as "tug of war"\cite{haertter2024stochastic,Ambarishkunwar2011mechanical}. We get this configuration when the PBB exhibits back and forth motion between the retrograde and the anterograde directions on the same microtubule. Here, it is understood that different sets of motors are pulling the cargo in two different directions. We observe this kind of a motion in the Fig. \ref{tugofwar}(a)-(b) and (d), where the cargo moves in the anterograde direction for a while and then suddenly turns around and moves back in the reverse direction. \textcolor{black}{We speculate that this motion could be on the same filament as the position and orientation has been better ascertained using the photonic force microscopy available with the optical tweezers.} Further, having a different filament exactly parallel to the first filament in such close proximity is unlikely. We find that this also leads to pitch rotational motion particularly at the point of turning (demonstrated in Fig. \ref{tugofwar}(i)) as reported earlier for the in-plane rotational degree of freedom\cite{ramaiya2017kinesin}. Our findings thus may constitute a novel reporting of pitch rotation during a tug-of-war event.

We show a 3-dimensional trajectory for a case which we putatively ascribe to the tug-of-war mechanism in Fig. \ref{tugofwar}. We call the motion between about 8 and 10 seconds to be on the same filament as the anterograde and retrograde tracks are just separated by about 20 nm, and indeed compatible with the same microtubule filament which is known to be about 40 nm wide. 

In Fig. \ref{tugofwar}f, we plot the mean square displacement (MSD) of the 
microparticle trapped inside the cell. The exponent of the (log)MSD is between one and two, and thus constitutes a signature of activity. To see this, we note that the medium is viscoelastic \cite{catala2025measuring}, and for passive motion, we would have expected a sub-diffusive motion (exponent less than one) \cite{yang2017application}. 
\textcolor{black}{Instead we find a positional MSD exponent larger than one, which indicates that the motion of the microparticle inside the cell is indeed active.} The pitch rotations, however, do not show clear deviations from a passive baseline. A useful way to quantify the net phase space (i.e the combined positional and rotational degrees of freedom) deviation from equilibrium is to measure the phase space currents. We do this for a minimal model that we construct in Section \ref{sec:model}; where we also provide explicit expressions for the phase space currents. In Fig. \ref{tugofwar}h, we plot this, and see clear circulating currents in the configuration space. The model is fitted to experimentally obtained pitch-$x$ \textcolor{black}{cross-correlation-function (CCF)} in Fig. \ref{tugofwar}g to obtain the parameters. We define this and explains its significance further in Section  \ref{sec:model} below, but we note here that
\textcolor{black}{these parameters when substituted into the expression of current - see Eq.\eqref{eq:current} - can be used to compute the current. 
The circulating current in our system is shown in Fig. \ref{tugofwar}h.
Thus, we find clear signatures of non-vanishing circulating currents, which
clearly demonstrates the broken detailed balance of the dynamical system \cite{muenker2024accessing}. 
}

\subsection*{Pitch rotation due to  cytoskeletal obstacle}

\begin{figure*}
    \centering
    \includegraphics[width= \linewidth]{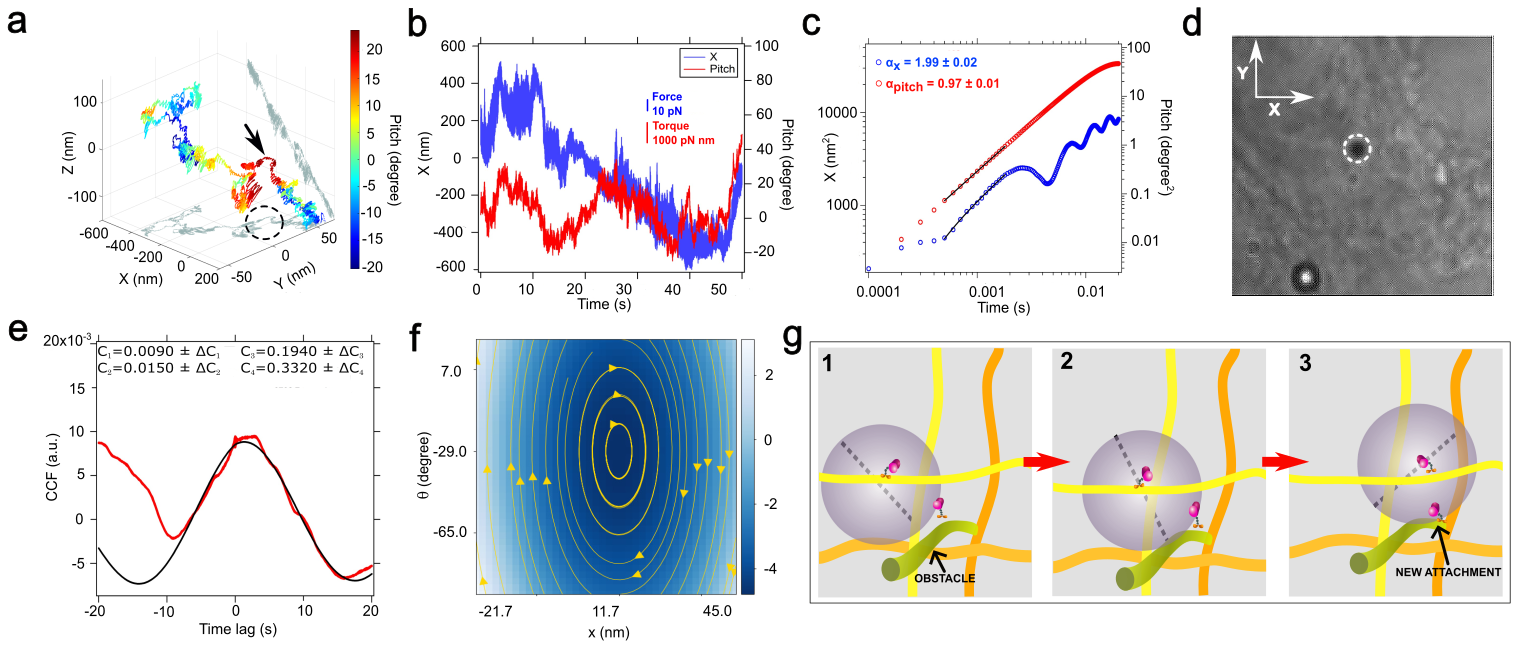}
    \caption{Trajectory of phagocytosed bead while encountering an obstacle (a) 4D trajectory, with the fourth dimension, pitch angle, shown as colour. The projections in XY and XZ plane are shown in grey. The sudden bump (marked with black arrow and within dashed circle in XY projection) indicate curving of trajectory around a possible obstacle which is accompanied by a large change in angle (in dark red).
    (b) Time series for simultaneous X (blue) and pitch (red) motion recorded with QPD with respective force and torque scale bars (c) MSD curve for X (blue open circle) and pitch (red open circle) with respective exponents shown as $\alpha_x$ and $\alpha_{pitch}$ indicating super diffusive and near diffusive behaviour respectively (d)bright field image of the cell showing position of the trapped phagocytosed bead enclosed in white dashed circle (e) CCF of pitch and X (in red) fitted to eq. (\ref{eq:CCF}) (in black line).
    (f) Using the fit to the data, we do a vector plot of the configuration current in phase space, see Eq.\eqref{eq:current}, which is overlaid on the pseudo-color plot of the steady-state probability distribution, see Eq.\eqref{eq:probS}. 
    (g) cartoon showing a spherical cargo dragged over an obstacle by motors (1-3) with black dashed line showing the direction of optic axis.  }
    \label{obstacle}
\end{figure*}

 In addition, we report events where the PBB under optical trap shows overall unidirectional motion. The emerging unidirectional nature of the PBB are observable when tracked over sufficiently long timescales of the order of 10 seconds. Fig. \ref{obstacle}(a)-(b) shows an event where the trajectory bypasses an obstacle during which the particle also turns in pitch sense. Microtubule intersections or overlying actin tracks pose an obstacle to the path of the cargo\cite{holzbaur2010coordination}. The motion is slower here than a free particle with motors due to the influence of the optical tweezers. \textcolor{black}{A similar kind of motion is also shown in Supplementary Fig.3.} 

The Fig. \ref{obstacle}(c)-(f) indicates that the trajectory indicated here is indeed active. 
In Fig. \ref{obstacle}c, we plot the MSD of the 
microparticle trapped inside the cell. Similar to the previous class of motion, we find the positional MSD we find to show signatures of activity. This is again corroborated by the phase space currents in Fig. \ref{obstacle}f. 
The rotations could be explained by a mechanism where the cargo has been attached at multiple positions (Fig. \ref{obstacle}(g)) inside the cell. Let us consider the case of a molecular motor attachment close to the equatorial plane of the cargo. While carrying the cargo, the particle suddenly experiences an obstacle either passive or active due to a secondary attachment in the polar region (either due to another molecular motor or other kinds), then it will turn. It is this rotation that we observe. Thus this is still a manifestation of activity in the system - these obstruction events can be thought of as externally imposed on the system, but the energy for the rotation would still come from the translational pull due to the motor. The cargo will be held at two locations, and thus, will turn. 

\textcolor{black} {It may also be noted here that we believe the point on the phagosome where the secondary attachment appears continues to remain attached until the obstacle is bypassed. If this secondary attachment point is not connected, the particle cannot experience a torque, since it takes two forces separated by a distance to generate it. It is remarkable that such dynamics can be observed for molecular motors which are such loosely attached to the phagosome shell in the azimuthal direction. }

\subsection*{Moving from horizontal direction to vertical segments}

\begin{figure*}
    \centering
    \includegraphics[width=\linewidth]{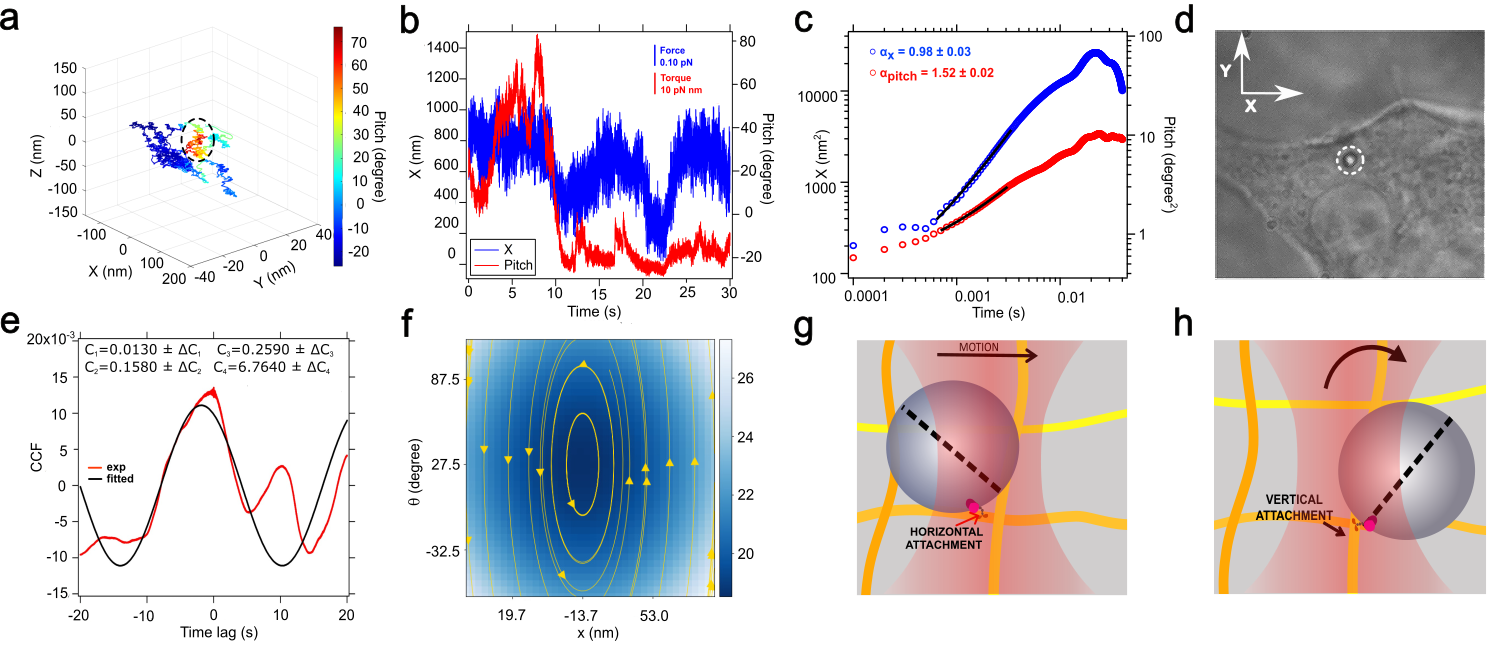}
    \caption{Change in plane of motion as point of attachment changes from horizontal to a vertical filament. (a) 4-dimensional trajectory of the PBB inside cell showing change in plane of trajectory from horizontal to vertical (shown within dashed ellipse) along with large change in pitch angle(in red) (b) X and pitch displacement recorded with QPD with respective force and torque scale bars (c) corresponding MSDs for pitch and X with exponents indicating near diffusive($\approx 1$) and super diffusive($>1$) behaviour respectively. 
    (d) Position of the phagocytosed bead under examination (enclosed within white dashed circle) as seen under bright field.
    (e) CCF of pitch and X fitted to Eq. (\ref{eq:CCF}). (f) A vector plot of the configuration current, see Eq.\eqref{eq:current}, which is overlaid on the pseudo-color plot of the steady-state probability distribution, see Eq.\eqref{eq:probS}. This plot clearly shows circulating current, a signature of broken detailed balance in the system.
    (g) and (h) Change in pitch angle when point of contact of PBB shifts from a horizontally placed filament to a vertically placed filament}
    \label{vertical}
\end{figure*}

We report another class of events where the particle exhibits three-dimensional motion inside the cell, marked by prominent vertical movements during brief intervals; this is displayed in Fig. \ref{vertical} \textcolor{black}{and Supplementary Fig. 4}. We also find that the particle is oriented in the vertical direction while moving vertically; see Fig. \ref{vertical}a. The result is interesting since the cargo is now spherical, much different from the rod-shaped particles used in the past\cite{chen2017characteristic,kaplan2018rotation,song2024deep}, where the complications due to anisotropic drag are not present. Thus, this can yield a mechanism for the correlation between translation and rotation. Indeed  the optical trap may also apply another stimulus that assists in the rotating the particle in the vertical segments. \\

With respect to the MSD curves (Fig. \ref{vertical}b), we note that the arrest of the motion stems from the non-linear effects of the applied tweezers. 
 Sometimes the time scales of the activity are such that the tweezers cannot compensate for it, while there are times when it can. Thus sometimes, the MSD exhibits diffusive behavior while at other times active behavior. It depends upon the nature of the process moving the particle both in translation and in rotation.

\subsection*{Rotation together with translation even for anisotropic particles not confined in tweezers}

\begin{figure*}
    \centering
    \includegraphics[width=\linewidth]{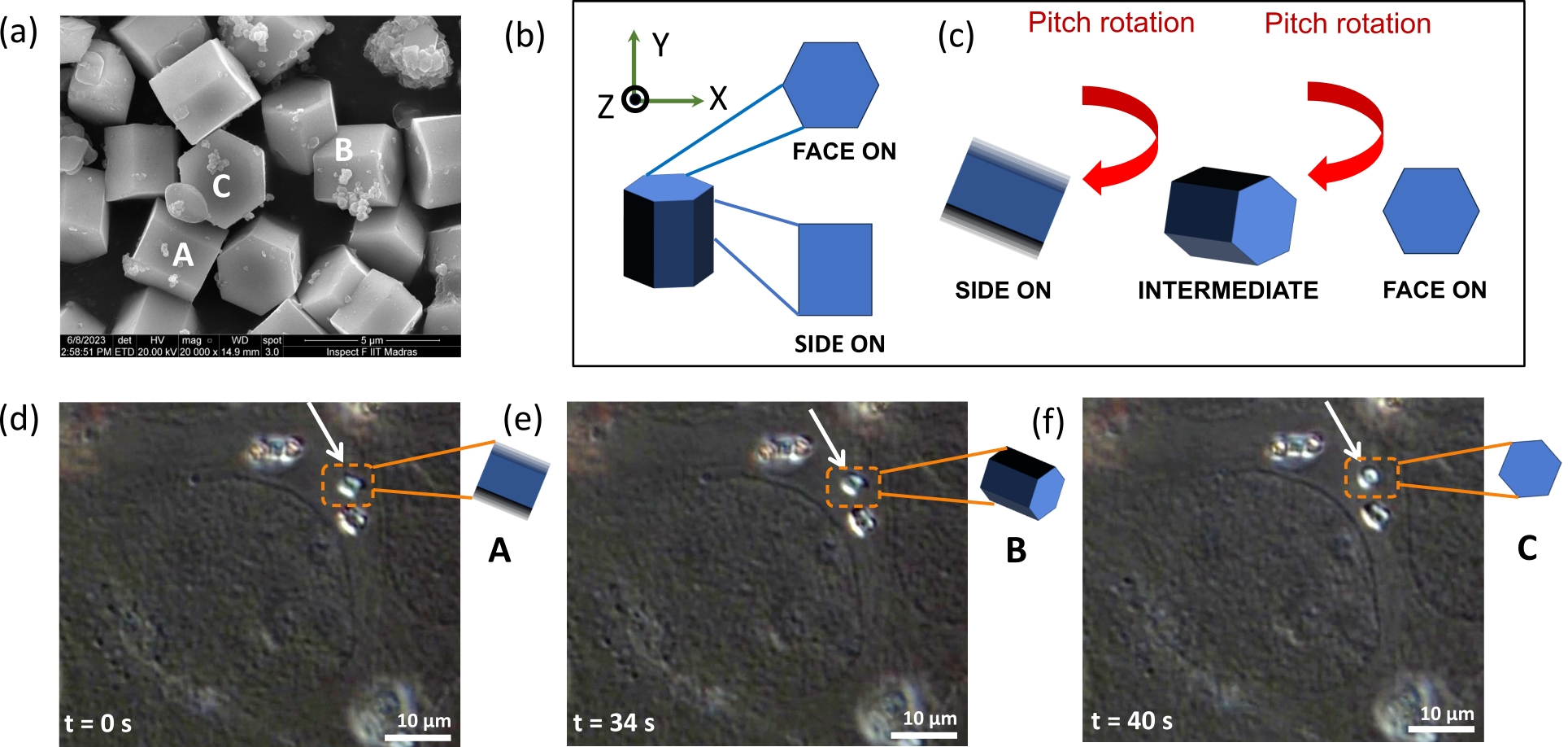}
 \caption{Complete rotation of phagocytosed hexagonal NaYF$_4$ crystals under intracellular activity in absence of optical trap (A) Scanning Electron Microscopy(SEM) images of NaYF$_4$ crystals with hexagonal prism shape (b) The hexagonal faces are called the 'face on' sides and the four rectangular faces are called 'side on' sides (c) When a crystal in 'side on'(A in (a)) orientation undergoes a pitch rotation, it orients 'face on'(C in (a))  via an intermediate orientation(B) in (a). (d)-(f)Phagocytosed NaYF$_4$ crystal inside MCF7 cell observed under bright field microscopy. Time lapse of the particle within dashed orange rectangle arrow shows a 90 degree flipping from (d) to (f). The particular crystal is seen to change orientation from A to C via B indicating rotation by 90 degree}
 \label{flipping}
\end{figure*}

To independently substantiate the presence of the observed rotational motion, we image the motion of isolated phagocytosed \textit{in vivo} hexagonal crystals ($\text{NaYF}_{4}$) via Phase Contrast Microscopy. Scanning electron microscope images reveals that these crystals have hexagonal face and rectangular sides in Fig. \ref{flipping}a (also shown in Fig. \ref{flipping}b). Change in orientation from `side on' to `face on' indicates a flip by 90 degrees in the pitch sense (Fig. \ref{flipping}c). This pitch rotation can be discerned during translation inside cells even in absence of an optical trap, as observed with phase contrast microscopy of live cells (Fig. \ref{flipping}d-f \textcolor{black}{(See Supplementary Movie 1)}). In particular, we confirm that we can get rotations, indeed sometimes about 90 degrees, in the pitch sense (Fig. \ref{flipping}(a),(e)). Thus, we conclude that the rotations inside the cell are induced by the internal active environment, independent of the optical trap.\\ 

\subsection*{Rotational and translational time series: Sudden jumps in the rotational motion uncorrelated with translation}

\begin{figure}
    \centering
    \includegraphics[width= 0.9\linewidth]{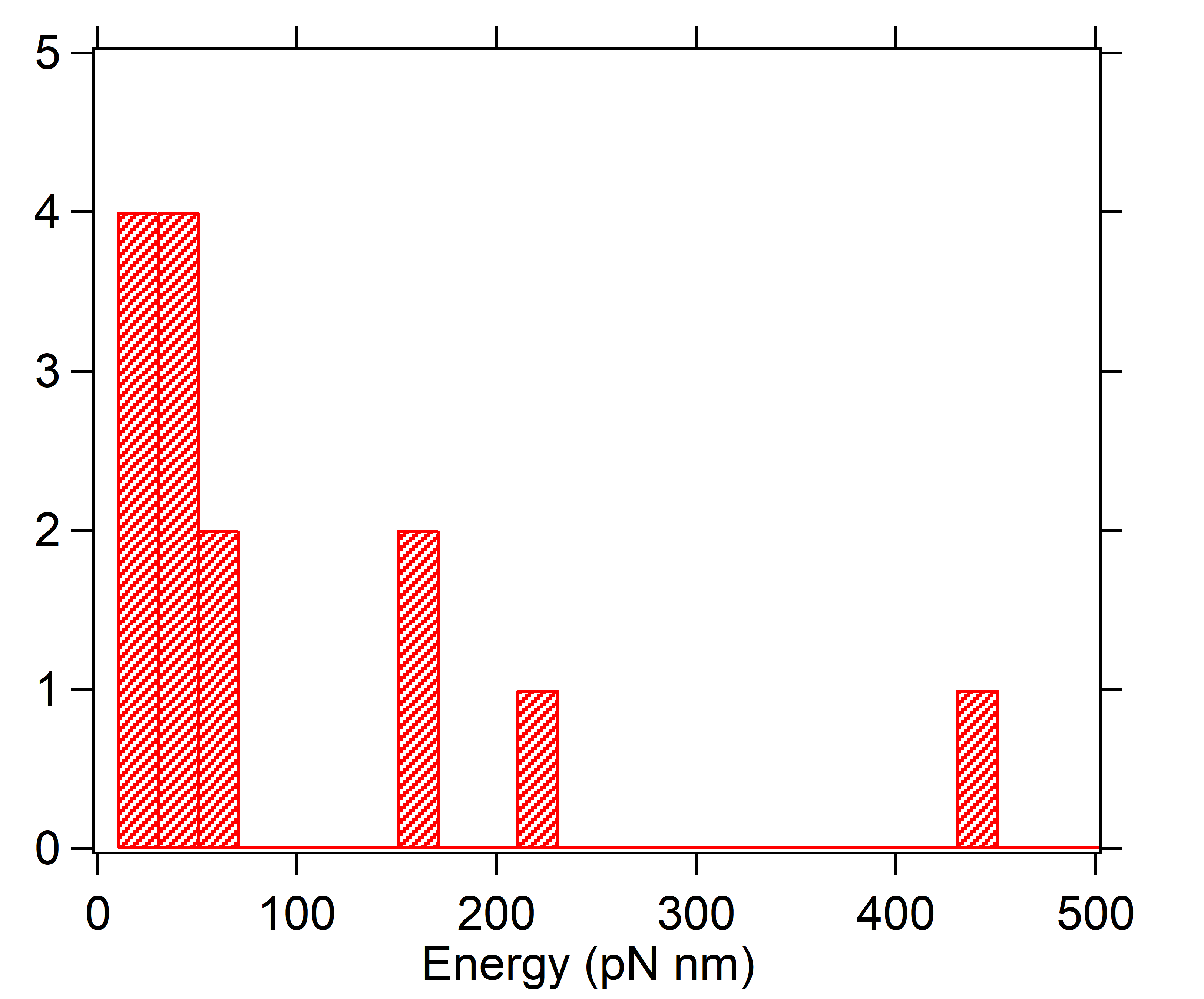}
    \caption{Statistics of rotational energy released when the phagosome slips with breaking of the obstacle.}
    \label{jumps}
\end{figure}

Beyond measurement of pitch angles, we can also infer the pitch torques  on the system. Thus, we are able to ascertain the "pitch energy" stored in the system. \textcolor{black}{Here, torque times the angle rotated gives the energy involved in the rotation.} We note that we also see sudden jumps in the pitch rotational time series which is hitherto uncorrelated with any corresponding jump in the translational time series. We plot the statistics of the energy released - see Fig. (\ref{jumps}) - in the sudden jumps and ascribe it to a possible breaking of an obstacle bond or an internal slip inside the PBB by the spherical particle causing the rotation. Given that the lipid bilayer of the phagosome is well attached to the internal particle, this kind of slip appears likely to be an attachment between the outer surface of the phagosome and an obstacle/secondary attachment point. 

If we consider that a molecular motor like kinesin stalls at a force of 2 to 3 pN \cite{block2007kinesin,blehm2013vivo,ramaiya2017kinesin}, and moves by 100 nm in this time, we can estimate that the motor builds up an energy of about 200-300 pN nm which is then released upon subsequent jump. Inside the cell, there may not be just one attachment point of the cargo to the filaments. One of the motors might still be carrying the cargo at an attachment point, but another secondary attachment point might yet break to release all of this energy which then manifests itself as the jump in pitch torque. Our rotational energy distribution in Fig. \ref{jumps} indeed indicate that we instead see a proliferation of ``low-energy slips", thus infrequent breaking of the molecular motor-filament bond. These low energy slips might be non molecular motor (or "passive") contacts. 

\textcolor{black}{The energy release of 200-300 pN nm while exhibiting sudden jumps in rotation is quite akin to sudden jumps of translation when the molecular motors like kinesin detach from microtubules and go back to the trap center. In such kinds of in-vitro kinesin motility assays, the motion of the kinesin stalls at about 5 pN after which, the tracer particle suddenly jumps back. Then the energy release is also similar, 5 pN $\times$ 100 nm. In our case of rotational jumps, the mechanism happens to be different. }

\begin{figure}
    \centering
  \includegraphics[width=0.9\linewidth]{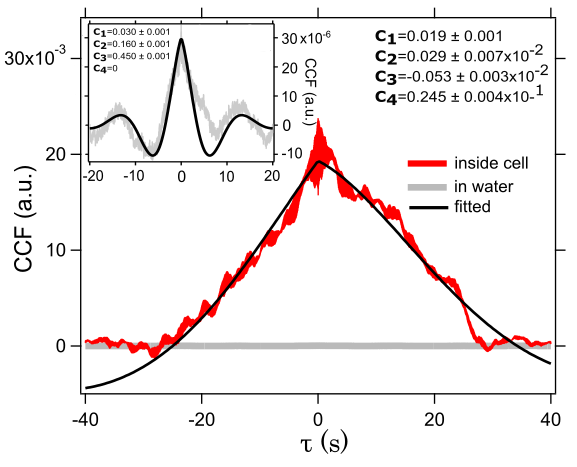}
    \caption{Comparison of CCF for a PBB inside a cell (in red) and a bead trapped in passive medium (water) shown in grey. Inset shows magnified CCF of bead trapped in water. Both are fitted with Eq. \ref{eq:CCF} with the fitted curve shown in black.}
    \label{waterCCF}
\end{figure}

\begin{figure*}
    \centering
    \includegraphics[width=0.95\textwidth]{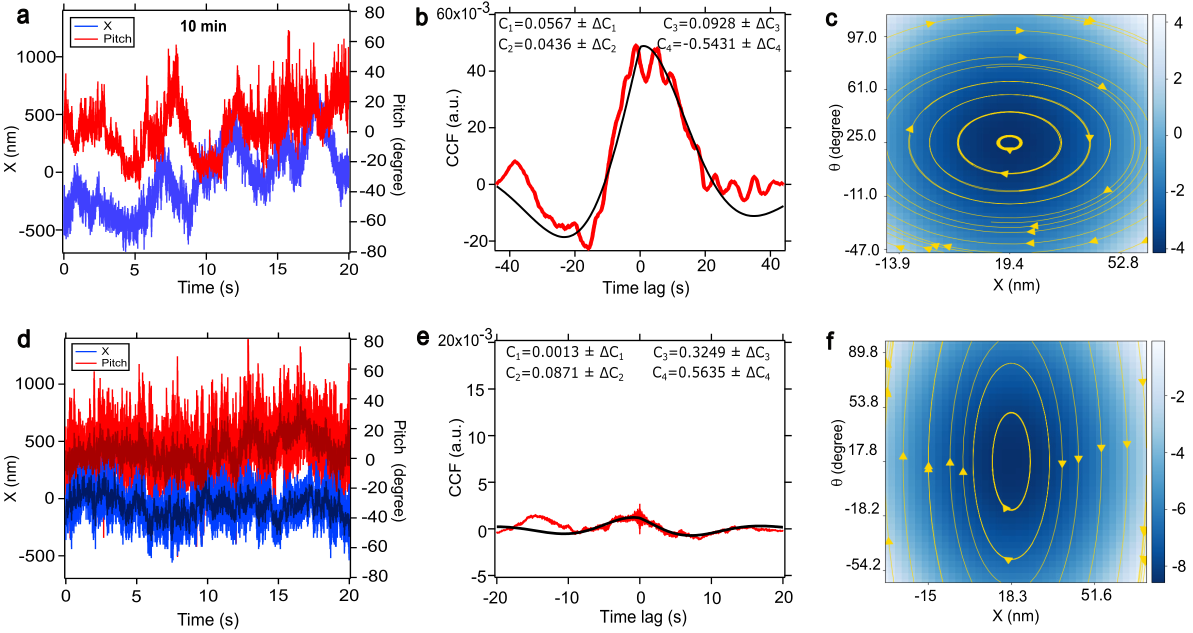}
    \caption{Motion of phagosomes in a cell at different times under suppressed activity. (a)  X (blue) and pitch (red) displacement time series (b) CCF between these two modes (c) vector plot of configuration current in phase space measured at 10 minutes after sodium azide addition. (d) and (e) are the displacement time series in X and pitch with same colour convention and their CCF respectively .(f) is the configuration current vectors in phase space mesaured at 30 minutes post addition of sodium azide. In (a) dark shaded curves represent the smoothened data.Experimental CCFs shown in red and black solid lines are the fit to it in (b) and (e)}
    \label{fig:sodium_azide}
\end{figure*}

%
\begin{table*}[t]
\caption{The parameters C$_{1}$, C$_2$, C$_3$, C$_4$ extracted by fitting the experimental data of CCFs to the theoretically predicted values from the minimal model given in Eq. \eqref{eq:CCF}. See text for the details of the fitting procedure.}\label{<table-label>}%
\begin{tabular}{@{}lllllllll@{}}
\toprule

Fig. &   C$_1$ & C$_2$   & C$_3$  & C$_4$   & $\Delta$ C$_1$   & $\Delta$C$_2$   & $\Delta$C$_3$       &  $\Delta$C$_4$\\
\midrule
2g     &  0.0070 & 0.0310  & 0.1200 & 0.4690  & 0.0001           & 0.00005         & 0.00003    & 0.0004\\
3e     &  0.0090 & 0.0150  & 0.1940 & 0.3320  & 0.00005 & 0.00009         & 0.00005    & 0.0004\\
4e     &  0.0130 & 0.1580  & 0.2590 & 6.7640  & 0.00003 & 0.0009          & 0.0001              & 0.0009\\
7      &  0.0190 & 0.0290 & -0.0530& 0.2450  & 0.0010           & 0.0001        & 0.00003             & 0.0004\\
8b     &  0.0567 & 0.0436  & 0.0928 & -0.5431 & 0.0001           & 0.0001          & 0.00001         & 0.0007\\
8d    &  0.0013 & 0.0871  & 0.3249 & 0.5635  & 0.000003 & 0.0006          & 0.0004              & 0.0017\\
10b    &  0.0160 & 0.0008& 0.0350& 0.2070  & 0.0010           & 0.00001          & 0.00006           & 0.0001\\
11b    &  0.0152 & -0.0380  & 0.1130 & 6.7880  & 0.0010       & 0.0090         & 0.0040             & 0.0060\\
13a    &  0.0107 & 0.0195  & 0.0739 & 0.1523  & 0.00001 & 0.0002         & 0.00006       & 0.0009\\
13b    &  0.0250 & 0.0100  & 0.0600 & 0.2362  & 0.0001           & 0.0001          & 0.0001           & 0.00001\\
13c    &  0.0357 & 0.0044  & 0.0582 & -1.8361 & 0.00002 & 0.00001         & 0.00001    & 0.0004\\
\botrule
\end{tabular}

\end{table*}

\subsection*{Correlated trajectories are indeed active ones}
\label{sec:model}

We have thus shown that the phagosomal translational and rotational modes are coupled. \textcolor{black}{Given that the phagosome's motion is driven by the presence of molecular motors moving on the cytoskeletal network}, we next ask whether this underlying non-equilibrium mechanism can be detected and/or quantified. In this section we will use a method that goes beyond the conventional MSD \cite{bacanu2023inferring, gnesotto2018broken} to show that this is indeed the case, and thus the non-equilibrium coupling here is in stark contrast to those encountered in purely diffusive systems \cite{volpe2006torque, roy2016using}. 

\subsection*{Minimal Model}\label{sec:theory}
\textcolor{black}{The above findings suggest two possible mechanisms of molecular motor attachment on the phagosome. The first is that of symmetric polarity, where motors of the same polarity attach on either side of the particle (e.g. kinesin-kinesin), causing a drift in translational motion. The second would be a differential attachment rate of anti-polarized motors (e.g. kinesin-dynein during tug-of-war). The simplest possible model to incorporate these two effects is:} 


\begin{align}
    \dot {\bm X}
    =-\bm A\cdot \bm X+
 \sqrt{2\bm D}\cdot \bm \xi
 \label{eq:mainLE}
\end{align}
Here, the position of the colloid is represented by $x$, while the angle made by it with the $\hat {\bm x}$-direction is given by $\theta$, with $\bm X = (x,\theta)$. Further, we have defined:
\begin{align}
    \bm A =
    \left(
    \begin{array}{cc}
        \mu^t \kappa_t\,\, & \mu \Omega_1 \\
     -\Omega_2/\mu & \mu^r \kappa_r
    \end{array}
    \right),\qquad 
    \bm D =
    \left(
    \begin{array}{cc}
        D_t & 0 \\
         0  & D_r
    \end{array}
    \right)
\end{align}
\cite{roy2016using}.\\ 

In the above, we have defined $\mu=\mu_t/\mu_r$, $D_t= {k_B T\mu^t}$ and $D_r= {k_B T\mu^r}$, where $k_B$ is the Boltzmann constant and $T$ is the temperature. 
Here $\mu^t=1/(6\pi\eta b)$ and $\mu^r=1/(8\pi\eta b^3)$ are, respectively,
the translational and rotational mobility of a spherical colloidal particle of radius $b$ in a fluid of viscosity $\eta$. 
Here,  $\cdot$ denotes maximal contraction between two tensors. 
The stochastic variable $\bm \xi=(\xi_1,\xi_2)$ are white noises, which have zero mean:
$\langle  \xi_i\rangle = 0$  and 
no correlation in time:
$\langle \xi_i (t) \xi_j (t')\rangle =\delta_{ij} \,\delta(t -t')$. 
\textcolor{black}{$\Omega_1$ and $\Omega_2$ we label as the \textit{symmetric attachment rate} and \textit{anti-symmetric attachment rate} parameters respectively. These two terms model two separate possible attachment types of molecular motors on the spherical particle. The model being sufficiently minimal, $\Omega_2$ can be also interpreted as, for instance, the net detachment rate of molecular motors upon encountering obstacles along its direction of motion (in obstacle avoidance). In either case, the net effect of motor attachment on either side breaks the symmetry and there is a net torque on the system. 
As shown below, the activity in our model arises from $\Omega_1$ and $\Omega_2$ terms.} In the limit of $\Omega_i=0$, our model corresponds to a Brownian particle at thermal equilibrium in a harmonic potential (equivalent to our bead placed in water - see Fig. \ref{waterCCF}). In the following, we fit the parameters of our model to identify the non-equilibrium features of the experimental system. 

The above model corresponds to a two-dimensional OUP (Ornstein-Uhlenbeck process) \cite{gardiner1985handbook}. This is an appropriate description as the colloids are trapped inside a laser beams which provides the harmonic confinement of the OUP \cite{volpe2006torque, roy2016using}. We write the model generally as a MVOUP (multi-variate Ornstein Uhlenbeck process) and specialize to the bi-variate case.We would like to note that we assume a linear model, which is motivated in parts by biophysical simplicity and analytical tractability. Adding nonlinearities, although numerically possible, would preclude analysis of non-equilibrium quantities (i.e EPR cannot be computed in closed analytical form). For the linear MVOUP model, closed form expressions for the EPR are available, as we show below.
In addition, the anti-symmetric coupling constants are chosen to demonstrate the non-equilibrium nature of the system. In particular, given that the tug-of-war dynamics are an important finding from experiments, such a coupling, including linear friction and stiffness coefficients, is an obvious starting point. Although a similar model has been used in non-biological systems \cite{volpe2006torque}, it has also been used in \textit{in vitro} colloidal systems \cite{roy2016using, ahmed2018active, nakul2023studying}. Accounting for more nuanced physical and/or biological details, such as hydrodynamic interactions, is beyond the scope of this work.

\subsection*{Experimental fit of time series}
 The correlation between the components of the MVOUP can be written as:
\begin{align}
    \mathcal C_{ij}(t)=\langle X_iX_j\rangle.
    \label{eq:corrXiXj}
\end{align} 
At the steady-state we obtain a  diagonal covariance matrix 
($ \bm c=\lim_{t\rightarrow \infty} \boldsymbol{\mathcal {C}}(t)$) be obtained analytically as \cite{singh2018fast,gardiner1985handbook}. It is given as:
\begin{align}
    \bm c =
    \left(
    \begin{array}{cc}
       k_B T/\kappa_t & 0 \\
         0  & k_B T/\kappa_r
    \end{array}
    \right)
\end{align}

In \eqref{eq:corrXiXj}, we have defined the general form of the correlation function of our model.  To compare the model with experimental data, we now consider the correlation function:
\begin{align}
    \langle
    x(t)\,\theta(t+\delta t)
    \rangle
     = \mathrm{C}_1e^{-\mathrm{C}_2|\delta t|}\cos (\mathrm{C}_3\delta t + \mathrm{C}_4)
     \label{eq:CCF}
\end{align}
Here $\mathrm{C}_1 = {k_B T}/{\kappa}$, $\kappa=\sqrt{\kappa_t \, \kappa_r}$, 
$2\mathrm{C}_2= \kappa_t \mu_t +  \kappa_r \mu_r$, and 
$2\mathrm{C}_3=\sqrt{4\Omega_1\Omega_2-(\mu_t \kappa_t-\mu_r\kappa_r)^2}$, $\mathrm{C}_4$ is a phase shift. 
\textcolor{black}{
$\mathrm{C}_1$ is the amplitude of the roto-translational correlation, $1/\mathrm{C}_2$ is the typical time scale at which these correlations decay, $\mathrm{C}_3$ is the frequency of the (oscillatory component of the) coupling, and $\mathrm{C}_4$ a phase shift.  Thus the parameter, $1/\mathrm{C}_2$, which is obtained by fitting to the experimental data quantifies the temporal correlations between the position of the colloid $x$, 
and the angle made by itself with the $\hat {\bm x}$-direction is given by $\theta$. 
We thus see that a stronger trap stiffness reduces roto-translational correlations, whilst enhanced coupling between molecular motors of the same and different polarities (non-reciprocal) is reflected in more sharply varying correlations within a given time interval.
The dimensions of parameters $\mathrm{C}_i$ are: $[\mathrm{C}_1]=[\text{L}]$, 
$[\mathrm{C}_2]=[\text{T}^{-1}]$, $[\mathrm{C}_3]=[\text{T}^{-1}]$, 
and $[\mathrm{C}_4]=[1]$, while $[\kappa]=[\mathrm{MLT}^{-2}]$. 
Throughout the paper, these constants have been provided in SI units and explicit units are not mentioned for due to paucity of space. We obtain these parameters by fitting to correlation functions computed on experimental data. \textcolor{black}{A description of the fitting procedure is available in the Methods section.} }

\subsection*{Non-equilibrium signature via probability flux}
Eq. (\ref{eq:mainLE}) describes a linear system in the phase space of $\{ x, \theta \}$ with an non-reciprocal evolution matrix $\mathbf{A}$ subject to white noise. For such class of systems, probability fluxes about the steady-state can be computed \cite{bacanu2023inferring, gnesotto2018broken,cates2022stochastic}, that characterize the out-of-equilibrium nature of the system.

To compute the non-equilibrium signature of the above model, we turn to the Fokker-Planck description.
The Fokker-Planck equation for the probability distribution $p(\mathbf{X}, t)$ - which correspond to Eq.\eqref{eq:mainLE} - is:
\begin{align}
    \frac{\partial p}{\partial t}=-\nabla\cdot\left[ -
    \mathbf A \cdot \bm X -\bm D\cdot \bm \nabla\right] p
    \equiv-\nabla\cdot\bm{J}.
\end{align}

The Fokker-Planck description for the above MVOUP can be solved to obtain the steady-state distribution \cite{singh2018fast,gardiner1985handbook}. The final expression is:
\begin{align}
p_{\mathrm{s}}(\mathbf{x})&=\left[2\pi{\sqrt{\operatorname*{det}\bm c}}\right]^{-1}\exp\left({-{\frac{1}{2}}\mathbf{X}^{T}\bm c^{-1}\mathbf{X}}\right),
\label{eq:probS}
\end{align}
The above expression of the probability can then be used to obtained the current using the Fokker-Planck description. The current in the system at the steady-state is given as:
\begin{align}
    {\bm J}_{\mathrm{s}}({\bm X})&=\left(\bm \Lambda\cdot \bm X\right) p_{\mathrm{s}}({\bm X}).
\label{eq:current}
\end{align}
The matrix $\bm \Lambda$ can be obtained as:
\begin{align}
    \bm \Lambda = -\bm A + \bm D \bm c^{-1} =-
    \left(
    \begin{array}{cc}
        0 & \mu \Omega_1 \\
     -\Omega_2/\mu & 0
    \end{array}
    \right).
    \end{align}
It is clear from the above that the matrix $\bm \Lambda$ vanishes if there is no roto-translational coupling (motor attachment to phagosome) in Eq.\eqref{eq:mainLE}, and thus, there is no current. 
By computing the above matrices, we can compute the current in the system once all the parameters of the model has been obtained by fitting the experimental data. Computing the curl of the current gives the condition for $\Omega_1$ and $\Omega_2$ for circulating current in the steady-state.

The steady-state flux is displayed for each of the types of motions reported - the tug-of-war in Fig. (\ref{tugofwar}(h)), the obstacle avoidance in Fig. (\ref{obstacle}(f)) and the vertical shifts in Fig. (\ref{vertical}(f)) - respectively. The vectorial fluxes are displayed with solid arrowed lines, overlaid with the steady-state probability distribution (scalar field). 
We find that all classes of motion are qualitatively similar in that they map circulating currents in the phase space of $\{ x, \theta \}$. \textcolor{black}{As we show below, this qualitative feature is a necessary, but not sufficient measure to quantify the non-equilibrium nature of the system.}\\

\subsection*{Quantifying broken detailed balance from EPR}
It is known the non-vanishing current in steady-state are signature of a system being away from equilibrium \cite{peliti2021stochastic}. Indeed, the current in configurations space of a system can be characterized by the EPR (entropy production rate), which measures how far is the system from equilibrium, and thus, quantifies the extend of broken detailed balance.
The  expression of the EPR ($\dot {\mathcal S}$) for our model is given as \cite{cates2022stochastic,peliti2021stochastic}:
\begin{align}
\dot {\mathcal S}_{\mathrm{s}}  = \int d\mathbf{X}\, \dot \sigma_{\mathrm{s}}(\mathbf X),\qquad 
\dot \sigma_{\mathrm{s}}(\mathbf X) =  \mathbf F^J\cdot {\bm J}_{\mathrm{s}}
\label{eq:EPR}
\end{align}
Here, we have defined a thermodynamic force $\mathbf F^J$, which is given in terms of the configurational current. It is given as \cite{BroeckEPR}: 
$\mathbf F^J=({\bm J}_s\cdot \bm D^{-1} )k_B/p_{\mathrm{s}}$. Thus, it measures if there is a non-vanishing current in the system.

\begin{figure}
    \centering
    \includegraphics[width=0.96\linewidth]{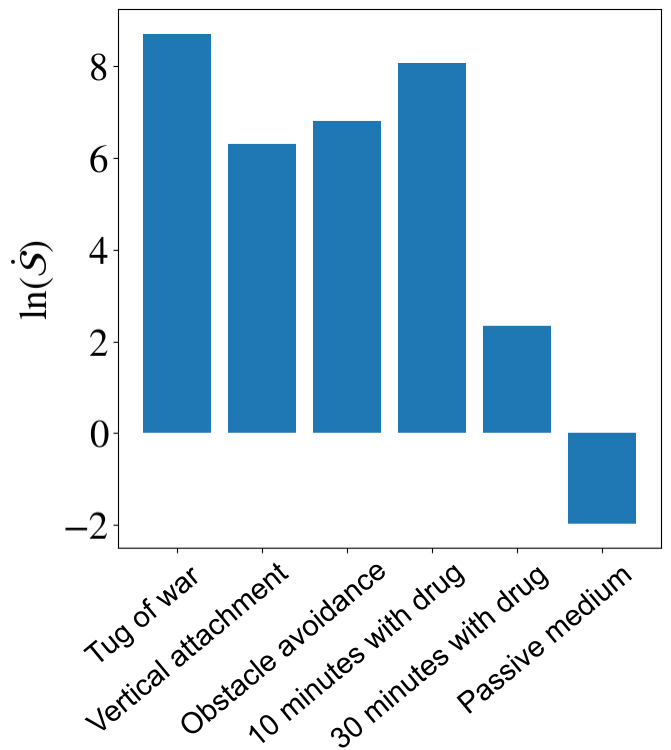}
    \caption{Logarithm of the EPR (entropy production rate) using Eq.\eqref{eq:EPR} from experimental data of three categories of trajectories described in the paper. \textcolor{black}{EPR at 10 minutes and 30 minutes are after adding activity suppressing drug to the cells (corresponding to Fig.\ref{fig:sodium_azide}(b) and (e) respectively.} EPR clearly demarcates the case of colloid in water (Fig. \ref{waterCCF}) when the the detailed balance condition is satisfied. Here EPR vanishes such that its log is negative.
    EPR is demonstrably larger for the case of phagocytosis. }
    \label{fig:EPR}
\end{figure}

The EPR (entropy production rate) 
of Eq.\eqref{eq:EPR} 
measures how far is the system away from equilibrium\cite{lynn2021broken,MichaelMurrell2024energy}. A vanishing 
EPR implies an effective Boltzmann description for the steady-state.
In addition to computing fluxes in the phase, we quantify the broken detailed balance by computing the EPR.
This is plotted for the various classes of motion observed here in Fig.\ref{fig:EPR}. \textcolor{black}{It is clear that there is substantial EPR when the colloid is inside the cell while the EPR vanishes when the colloid is inside water. It is also evident that the TOW dynamics has the highest EPR of all the motion types observed. Upon addition of the surpressant, EPR is also substantially reduced relative to the control case. In addition, it is interesting to note that qualitative (circular) configuration currents for ``passive" systems studied here (e.g. Fig. \ref{tugofwar}(h), Fig. \ref{vertical}(f), Fig. \ref{obstacle}(f), and Fig. \ref{fig:sodium_azide}(c)) do not by themselves provide a measure of activity in the system; one indeed requires the EPR to make this distinction.} Overall, these results further substantiate the active nature of the coupling of the phagosomal motion in the cytoskeletal complex. \\

\subsection*{Rotation and translation after suppression of activity}

We attempt to arrest activity inside the cell by depleting ATP. The cells are treated with 2mM sodium azide (NaN$_3$) solution, known for  hindering oxidative phosphorylation of mitochondria and hence inhibits ATP generation \cite{guo2014probing}. The motion of phagosomes in ATP depleted cells are studied at different times. X and pitch time series recorded after 10 and 30 minutes from treatment of the drug shows significant decrease in motion as shown in Fig. \ref{fig:sodium_azide}(a) and (c) respectively \textcolor{black}{(See Supplementary movie 1)}. Amplitude of respective CCF curves (C$_1$) is high at earlier times (in Fig. \ref{fig:sodium_azide}(b) and reduces in Fig. \ref{fig:sodium_azide}(e) about 30 minutes later with increase in (C$_2$) to about twice of that in Fig. \ref{fig:sodium_azide}(b). Increase in (C$_2$) implies reduced correlation time and reduced correlation between the two modes of motion. Thus activity depletion directly affects the motion and correlation which further validates molecular motor activity as the source of correlated motion. The circulating configuration current vectors in phase space (Fig. \ref{fig:sodium_azide}(c) and (f)) show that the non equillibrium nature persists though with a reduced extent in Fig. \ref{fig:sodium_azide}(f) confirming that the cells are alive in presence of the suppressant. \textcolor{black}{Indeed it appears that the activity of the molecular motors happens in low frequency or long time scales, like slower than 100 msec, as is also noticable in Fig. \ref{fig:sodium_azide}(a) and (d).}

\section*{Discussions\label{sec:discuss}}

Intracellular motion comprises of a multitude of influencing factors. Rotational behaviour of spherical probes along with 3D tracking in the presence of a confining trap both in translation and in rotation elucidates the local cellular microenvironment with millidegree resolution. The roto-translational correlations and MSDs quantify the nature of phagosomal motion in vivo. Finite roto-translational correlations that we report are of an underlying non-equilibrium nature, initiated by the molecular motors while moving the cargo. The active rotational torque is then transferred to the inner spherical particle which turns in spite of the restoring torque applied by the optical tweezers, implying significant rigidity of the phagosome shell. We have also reported that, ocassionally, the torque built up in the system is suddenly released, either as a rotational slip inside the phagosome or as a slip from an obstacle outside. \\

We note that thermally induced Brownian motion is purely stochastic both in translation and rotation, thus does not amount to correlated motion. This does not generate torque to cause directed rotation, especially in presence of the linearly polarized optical tweezers trying to orient the particle along the direction of polarization. 


Indeed the point of contact of the particle to the cell membrane when getting phagocytosed at the cell boundary is known to be maintained even while inside the cell \cite{bohdanowicz2013role,swanson2008shaping}. There may be bonds formed between the particle and the phagocytic membrane, which are retained even later, implying that the lipid membrane is well held to the inner particle. Thus, it is likely that the slip events we noticed in Fig.\ref{jumps} occurs on the outer side of the phagosome. 

\textcolor{black}{We would also like to reiterate that the three classes of the events shown in the manuscript have been classified by eye estimation of the three-dimensional trajectories. Events where obstacles have been encountered have been identified by the presence of curved detour trajectories as opposed to the straight lines. In the events with vertical motion segments, these events are instead identified by sustained deviations from the x-y plane, etc. 
In all these cases, the main message of the manuscript hold good, namely there is pitch rotational motion in conjugation with translational motion.}


We note that it has been recently suggested that even 
"pacman" kind of motors can also cause intracellular 
activity \cite{umeda2023activity}. 
\textcolor{black}{However, we believe that it is unlikely that such motors can also apply torque to cause directed rotation.} Thus, it does appear that the intracellular activity is mainly due to processive molecular motor activity.

\textcolor{black}{Recent work by Vorselen et al. have shown large forces developed by the actomyosin cortex 
during phagocytosis \cite{vorselen2021phagocytic}. In another recent study \cite{poirier2020f} observed F-actin flashes on macrophage phagosomes that 
deforms their content, thereby facilitating digestion. These effects would assist in the hardening of the phagosome shell. Moreover, the shell could actually also comprise cytoskeletal fibers in addition to being fluidic,} which would make them rigid. The net affect of all this seems to make the shell rigid enough to transfer torque from outside to inside.

Intracellular calibration of optical tweezers is a very difficult task, conventionally believed to be complicated by unknown local refractive index variations. The optical trap stiffness varies from location to location inside a cell, while the viscosity also changes. These changes render the optical trap stiffness inside the cell variable for the same value of the optical tweezers laser power. Indeed this is the problem due to which such tweezers has been difficult to calibrate inside the cell. The community has used many different techniques like mimicking the cytosol with a fluid comparable in refractive index \cite{rai2013molecular}, or even not trying to directly measure the position, but rather estimating the forces directly with changes in momentum of the light \cite{RitschMarte2021direct,RitschMarte2022generally}. We use a strategy that we ourselves have developed for calibration. We assume that every trajectory inside the cell is unique and hence the fits to the power spectral density of that fit yields the local trap stiffness. Moreover, the same kinds of considerations also apply to the torques, and angular displacements with an additional facet for rotational trap stiffness being the birefringence of the particle used. This depends upon size and even the actual particle used. Thus, all the trap stiffnesses for the same value of optical tweezers laser power may not be the same.
Moreover, the noise band is a very complicated issue. As can be seen in Fig. \ref{vertical}b, the noise band initially between 20 to 25 sec is about 80 nm but beyond like 30 sec, it rises suddenly to 150nm. We believe there are many factors behind the increase of noise band, namely the health of the cell, internal activity and several other factors that we do not fully understand at the moment. Further, the viscosity of the cell changes not only from location to location but also temporally, making this kind of increase on noise a good topic of future research, but beyond the scope of the present manuscript.\\

\textcolor{black}{Intracellular calibration of optical tweezers is a very difficult task, conventionally believed to be complicated by unknown local refractive index variations. The optical trap stiffness varies from location to location inside the cell, while the viscosity also changes. These variations render the intracellular optical trap stiffness variable for the same value of the optical tweezers power.  The community has used various techniques like mimicking the cytosol with a fluid comparable in refractive index \textcolor{black}{\cite{balint2013correlative}}, or even not trying to directly measure the position, but rather estimating the forces directly with changes in momentum of the light \textcolor{black}{\cite{volpe2006torque}}. We use a strategy that we ourselves have developed for calibration. We assume that every trajectory inside the cell is unique and hence the fits to the power spectral density of that fit yields the local trap stiffness. Moreover, the same kinds of considerations also apply to the torques, and angular displacements with an additional facet for rotational trap stiffness being the birefringence of the particle used. This depends upon size and even the actual particle used. Thus, all the trap stiffnesses for the same value of optical tweezers laser power may not be the same.} 

\textcolor{black}{Moreover, the noise band seems to be a very complicated issue. As can be seen in Fig. \ref{obstacle}b, the noise band initially between 20 to 25 sec is about 30 nm but beyond like 30 sec, it rises suddenly. We believe there are many factors behind the increase of noise band, namely the health of the cell, and factors we do not fully understand at the moment. Further, the viscosity of the cell changes not only from location to location but also temporally, making this kind of increase of noise a good topic of future research , but beyond the scope of the present manuscript.}

To conclude, we report that a spherical particle inside a phagosome senses an (active) torque, transferred from molecular motors on the outer surface of the phagosome. We have further corroborated the non-equilibrium nature of this torque transfer by computing phase space fluxes and steady-state entropy production on a minimal biophysical model, which is significantly enhanced relative to passive baselines.\\

We envisage that the kinds of systems studied here can be used to study transport dynamics; for instance, to study mechanisms leading to roto-translational correlations. Moreover, this opens up the entire field of rotational rheology to assist in translational studies towards problems of the kind relevant to cell biology. As stated recently \cite{nakul2023studying}, the intracellular rheology tracking both activity and viscoelasticity can be used to study problems like chemical kinetics, activity enhanced diffusion and even how the cells maintain the internal heterogeneity with active processes.

%

\section*{Methods}

\subsection*{Optical tweezers setup}

The experiments are carried out in inverted microscopes integrated with optical tweezers setups; namely OTKB/M from Thorlabs and Ti2 Eclipse from Nikon. The schematic of the setup is shown in Fig.\ref{schematic}.(b). 1064nm beam from the laser source is passed through a polarising beam splitter(PBS). This polarised laser beam reflects off a dichroic mirror inclined at 45\textdegree towards the objective lens. We used an oil immersion, high numerical aperture(100X, 1.3NA, Nikon) objective which, besides tightly focusing, the beam at the sample chamber, also serves for high-magnification imaging. At the sample plane, the highly focused IR beam can trap micron sized particles. The forward scattered light along with unscattered fraction of the beam passes onto the condenser.  The outcoming light is reflected by another dichroic and passed through another PBS. This mixture of forward scattered and unscattered polarised light is made to pass through the PBS to split into orthogonal components which are then made incident on the photodetectors.

\subsection*{Preparation of liquid crystalline microspheres}

Liquid crystal microspheres having intrinsic birefringence are used for probing rotations. It is prepared using commercially available precursor RM-257 powder (2-Methyl-1,4-phenylene bis(4-(3-(acryloyloxy)propoxy)benzoate)). The precursor is available in amorphous form at room temperature.\\ 
Ethanol and water are taken in the ratio of 1:3 by volume in separate beakers and heated. The precursor powder is first added to ethanol heated to 55 C while being constantly stirred by a magnetic stirrer. The powder dissolves completely to form a transparent solution and is heated to 65 C allowing the precursor to undergo phase transition to nematic liquid crystal. This mixture is then added dropwise to water maintained at 75 C. The mixture is constantly stirred with a magnetic stirrer at 100 rpm. The solution turns milky white indicating the formation of birefringent spheres. The ethanol is then made to evaporate gradually, leaving behind a colloidal dispersion of nematic liquid crystal spheres (diameter depending on initial amount of precursor powder, ethanol to water ratio and evaporation rate) in water.

\subsection*{Preparation of living cells and insertion of microparticles}

The experiments were performed on breast cancer cells (MCF-7). Cells were initially maintained in Dulbecco's modified Eagle medium (DMEM) at 37 C with 5\% CO$_2$ in an incubator. Then the glass coverslips are cleaned with detergent and then by acid wash using 1M HCl. Thereafter, the cells are pelleted using a centrifuge and then resuspened in DMEM with an approximate concentration $\sim 10^3$ cells/ml. A drop of this cell suspension is placed at the centre of the cleaned coverslip and placed in the incubator and allowed to grow for 2-3 days till it reaches $50\%$ confluency. Sufficient amount of fresh media mixed with a concentrated suspension of $1 \mu m$ sized birefringent microspheres are then added to the petridish containing the seeded coverslip. It is incubated at 37 C with 5\% CO$_2$ for 24 hours till when the cells have ingested the microparticles via phagocytosis.

\subsection*{Experiment statistics}

\textcolor{black}{These experiments were performed in three months of a year. In total, there were about 10 days during which the experimental measurements were taken. Every day, about 2 to 3 cells were studied within the first 20 minutes of placing the sample on the setup to reduce chances of causing stress on the cell. In total about 50 trajectories were recorded out of which 35 showed significant rotation in the pitch sense. These 50 already excluded problematic events or experimental imperfections.}

\subsection*{Activity suppression in cells for control experiment}
\textcolor{black}{As a control experiment, motion of phagosomes were studied in ATP depleted cells to suppress the activity without killing the cell. To arrest molecular motor activity 2mM Sodium azide (NaN$_3$) solution was added with cell media. Cells were maintained under 37 C in sodium azide containing media for 5 min before being studied.}  

\subsection*{The experiment}
 Cells maintained in DMEM media are placed in the sample chamber as shown in Fig. \ref{schematic}(a). The sample chamber is constructed by placing another glass coverlip on the coverslip containing MCF-7 cells. Then it is placed between the condenser and objective, on the sample stage. The sample chamber is placed in inverted position such that the cells are attached to the upper surface of the sample chamber(Fig. \ref{schematic}. (b)). A polarized 1064nm laser beam is focused on the sample chamber through the objective. The sample stage can be moved in X,Y and Z directions. Particles inside the living cell are brought into focus of the laser to trap it. The laser power is maintained at about 50mW at the sample plane. Forward scattered light from the sample plane is collected by the condenser and split into two orthogonal components by a polarising beam splitter. Forward scattered intensity of the light is detected by the quadrant photodiode(QPD). A half wave plate (HWP) is placed before the PBS to ensure that light falling on QPD has the same polarization as the incoming beam. For pitch detection, a method similar to \cite{roy2023comparison,chakraborty2023high} is followed.  The orthogonal polarization is split into two equal halves using an edge mirror and sent to two ports of a balanced photodiode(BPD) which finds the difference in intensity between the two halves. This intensity difference is proportional to out of plane rotation of a particle in pitch sense. QPD and BPD is connected to a Data Acquisition system(NI, USA) via a monitor. Data is acquired at a sampling rate of 40kHz.

 \textcolor{black}{ The cells were placed upside down in the experiment because of lesser aberrations introduced in the polarization path. Of course, placement on the bottom surface also works, but we find this present configuration better for our pitch detection system. The sample chamber has buffer fluid in the region between the glass surfaces and is about 30-40 $\mu$m thick, quite consistent with a 160 $\mu$m glass cover slip at the bottom. The 1.3 NA objective can reach the top surface well.}

\subsection*{Calibration}
Calibration of the photonic force microscopy inside the cell has been complicated by the unknown nature of the refractive index of the cytoplasm, and even the subsequent variation with location and time inside the cell. It is here that we explore a technique recently suggested to take the power spectral density of the position and orientation fluctuations of a cargo inside a cell,and then extract calibration factors. A typical set of time series has been shown in Fig.\ref{schematic} (d). 

 We use passive calibration method as described in \cite{nakul2023studying} and \cite{vaippullyviscoelastic}. Stokes Oldroyd-B model for frequency dependent viscoelasticity (eq. (\ref{Oldroyd}))is used to define motion of a symmetric particle within an optically confined region in viscoelastic medium (as derived in \cite{paul2018free}.

\begin{equation}
    \mu(\omega)=\left[\mu_s+\dfrac{\mu_p}{1-i\omega\lambda}\right]
    \label{Oldroyd}
\end{equation}

This model is shown to yield the generalised Maxwell's model (or the Jeffery's model) for viscoelasticity. The situation is similar to overdamped harmonic oscillator whose power spectral density is given by eq. (\ref{PSD}). The rotational mode is modeled as a rotational spring and dashpot in the viscoelastic medium.

\begin{equation}
   \mathrm{ \langle x(\omega)x^*(\omega)\rangle}=\frac{\frac{2k_BT}{\gamma_0}\left(\dfrac{ 1+\dfrac{\mu_p}{\mu_s}}{\lambda^2} +\omega^2\right)}{{ {\left(\dfrac{\kappa}{\gamma \lambda}-\omega^2\right)^2+\omega^2 \left(\dfrac{\kappa}{\gamma}+\dfrac{1+\dfrac{\mu_p}{\mu_s}}{\lambda}\right)^2} }}+ y_0 
   \label{PSD}
\end{equation}

Comparing eq. (\ref{PSD}) to the generalised form given 
\begin{align}
   \mathrm{ \langle x(\omega)x^*(\omega)\rangle}&=\beta^2A\frac{\left(\dfrac{ 1+\dfrac{\mu_p}{\mu_s}}{\lambda^2} +\omega^2\right)}{{ {\left(\dfrac{\kappa}{\gamma \lambda}-\omega^2\right)^2+\omega^2 \left(\dfrac{\kappa}{\gamma}+\dfrac{1+\dfrac{\mu_p}{\mu_s}}{\lambda}\right)^2} }}\nonumber\\
   &+ y_0 \hspace{0.2cm}V^2/Hz
   \label{Volt_PSD}
\end{align}

Comparing eq.(\ref{PSD}) with eq. (\ref{Volt_PSD}), the calibration factor $\beta$ is calculated as $\beta=\sqrt{\frac{2k_BT}{\gamma A}}$ $m/Volt$ (or $rad/Volt$) for translation (or rotation). $\gamma_0$ accounts for viscous drag at 0Hz exerted by the medium. Considering cells as water based systems, $\gamma_0=6\pi\eta a$ for translational motion and $\gamma_0=8\pi\eta a^3$ for the rotational systems.

\section*{Acknowledgment}
We thank Erik Sch{\"a}ffer and Timo Betz for useful discussions. We also thank Saumendra Bajpai for providing us with MCF-7 cells and his valuable inputs for the control experiment with suppressed activity.
We thank the Indian Institute of Technology, Madras, India for their seed and initiation grants to BR and RS. This work was also supported in parts by the DBT/Wellcome Trust India Alliance Fellowship IA/I/20/1/504900 awarded to BR. 

\section*{Competing Interest}
The authors declare no competing interest.

\section*{Data Availability Statement}

The data with the manuscript are available from the corresponding author upon reasonable request. Some videos have already been furnished in the supplementary information.

\clearpage
\onecolumngrid   
\newpage
\begin{center}
    \textbf{\LARGE Supplementary Information}\\[2ex]
    
\end{center}

\begin{figure}[p]
\centering
\includegraphics[page=1,width=\paperwidth,height=\paperheight,keepaspectratio]{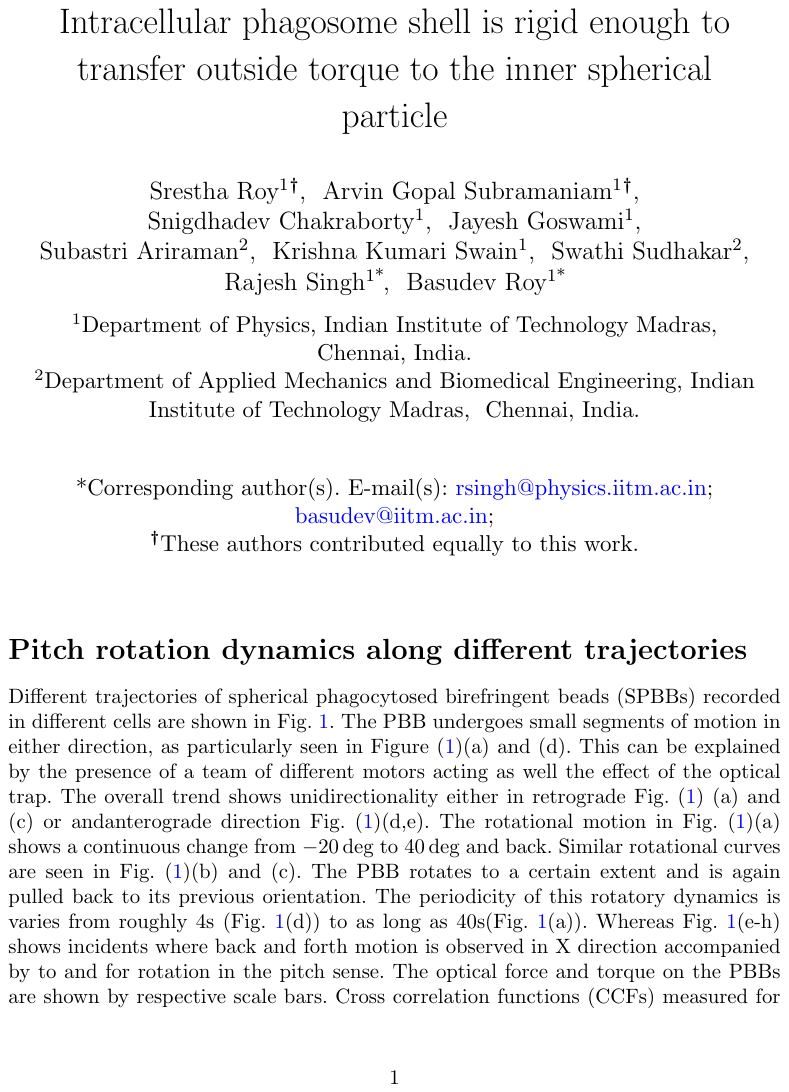}
\end{figure}

\clearpage
\begin{figure}[p]
\centering
\includegraphics[page=2,width=\paperwidth,height=\paperheight,keepaspectratio]{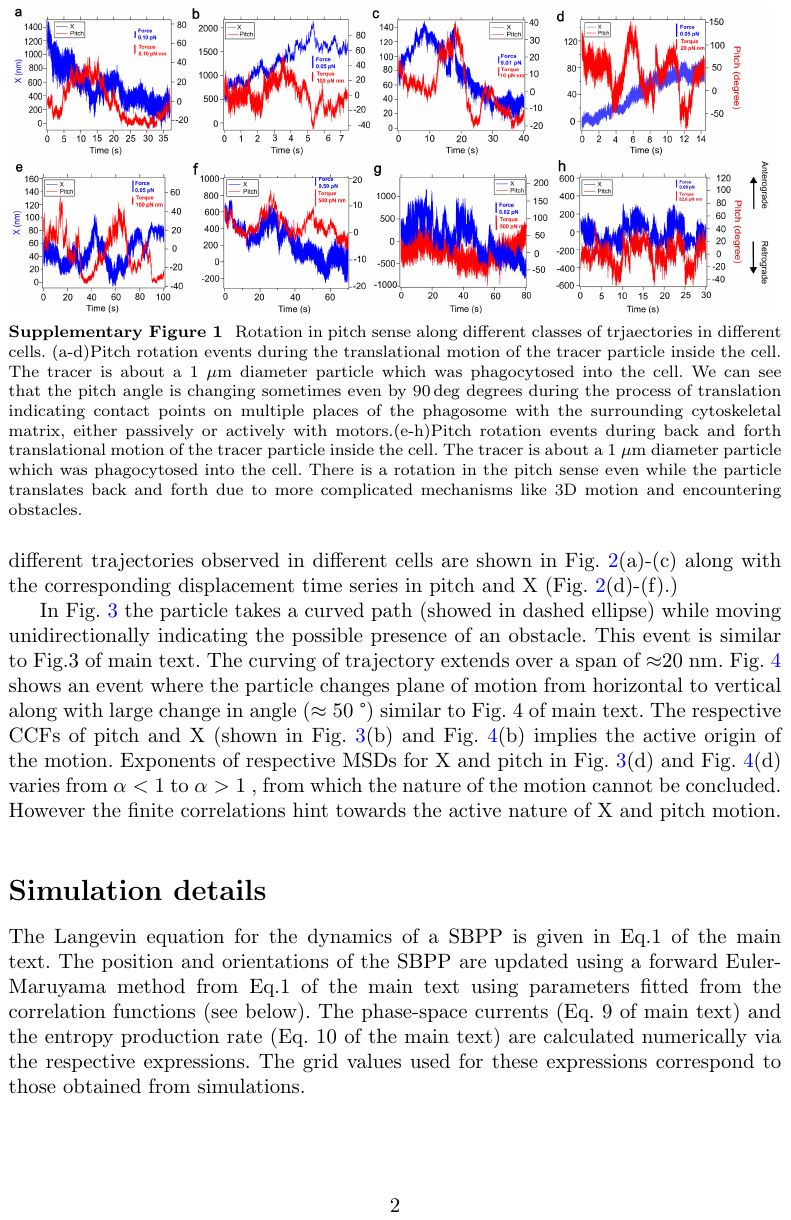}
\end{figure}

\clearpage
\begin{figure}[p]
\centering
\includegraphics[page=3,width=\paperwidth,height=\paperheight,keepaspectratio]{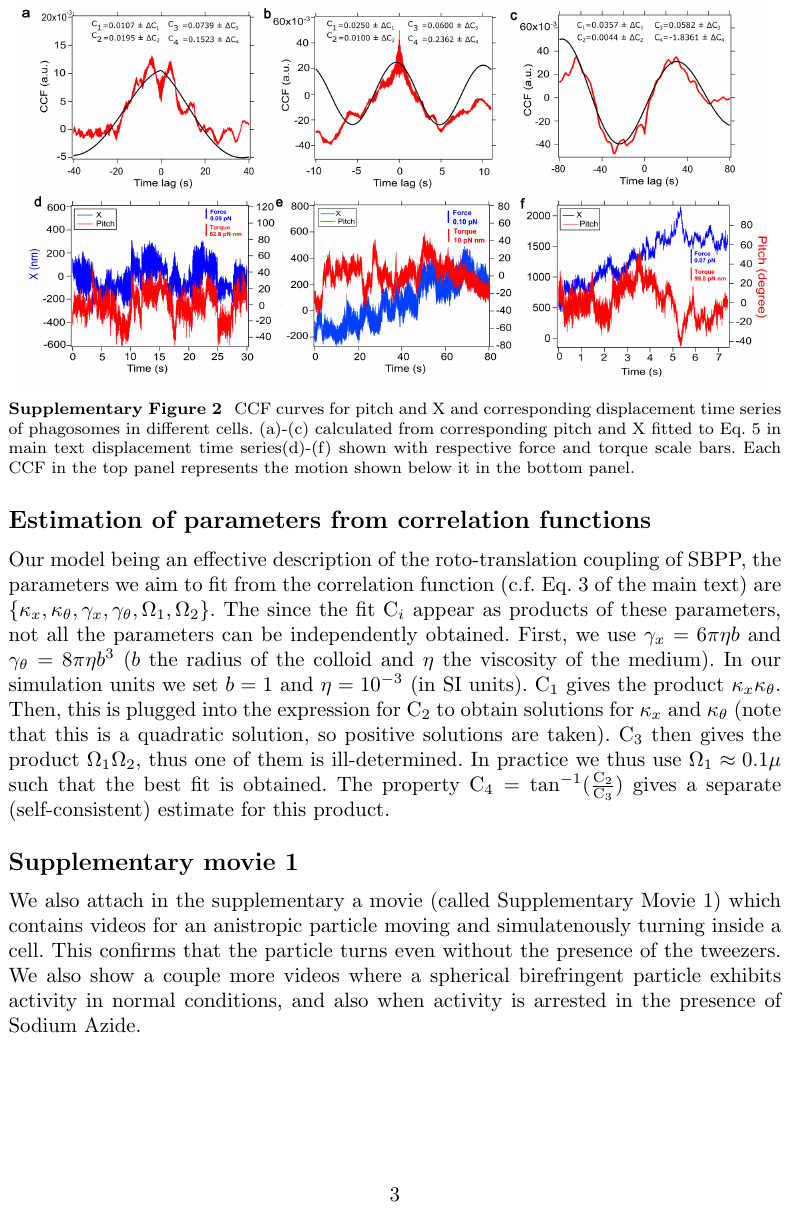}
\end{figure}

\clearpage
\begin{figure}[p]
\centering
\includegraphics[page=4,width=\paperwidth,height=\paperheight,keepaspectratio]{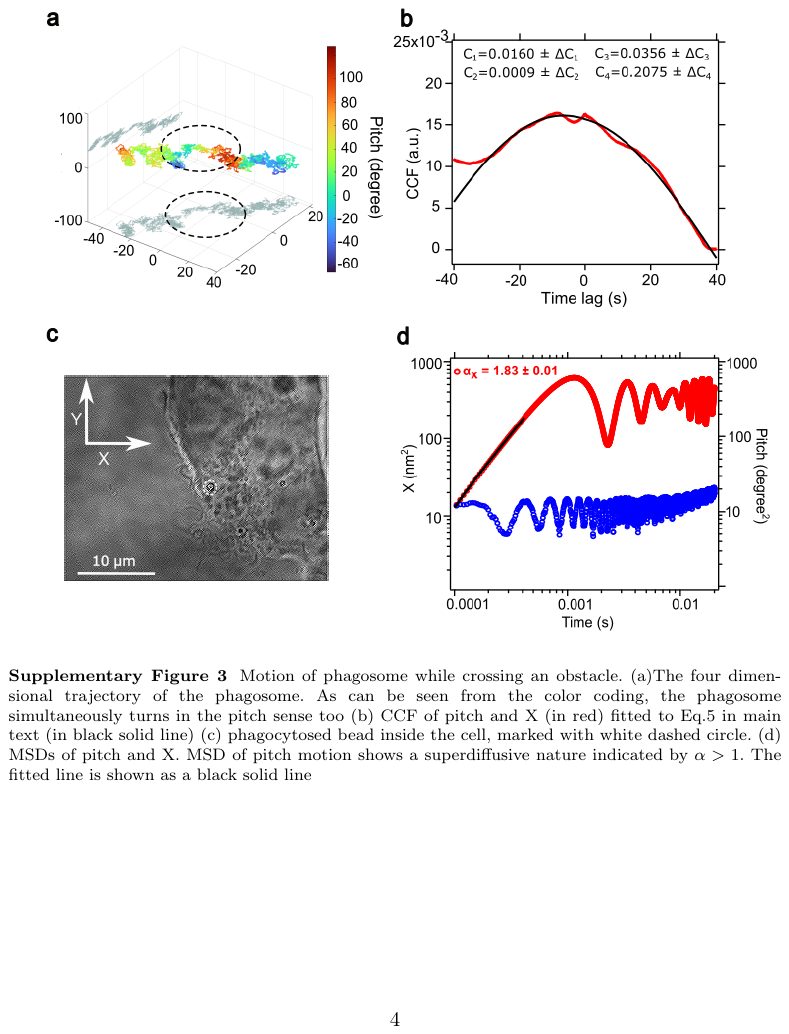}
\end{figure}

\clearpage
\begin{figure}[p]
\centering
\includegraphics[page=5,width=\paperwidth,height=\paperheight,keepaspectratio]{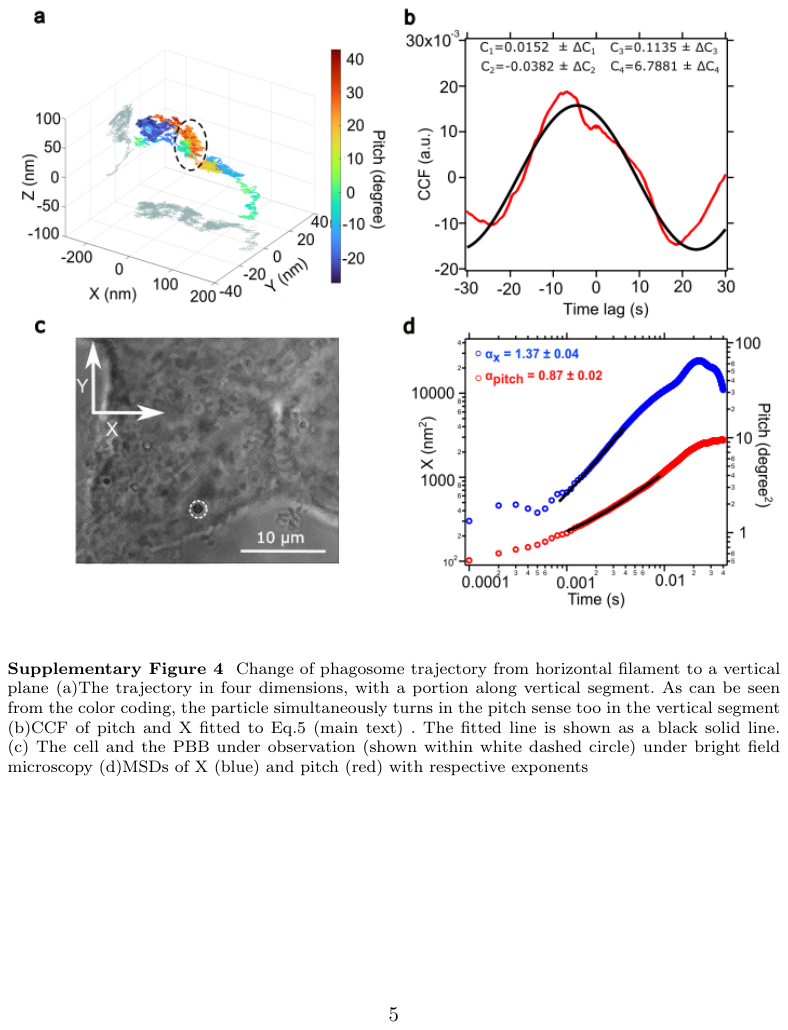}
\end{figure}

\clearpage
\twocolumngrid  

%
%


\end{document}